\documentclass[a4paper,
               keeplastbox,   
               ]{jacow}
\usepackage{pdfpages,multirow,ragged2e} %

\makeatletter%
	\ifboolexpr{bool{xetex}}
	 {\renewcommand{\Gin@extensions}{.pdf,%
	                    .png,.jpg,.bmp,.pict,.tif,.psd,.mac,.sga,.tga,.gif,%
	                    .eps,.ps,%
	                    }}{}
\makeatother

\ifboolexpr{bool{xetex} or bool{luatex}} 
 {}                                      
 {\usepackage[utf8]{inputenc}}           

\usepackage[USenglish]{babel}

\ifboolexpr{bool{jacowbiblatex}}%
 {%
  \addbibresource{jacow-test.bib}
  \addbibresource{biblatex-examples.bib}
 }{}
\begin{document}

\title{Dual frequency master oscillator generation and distribution for ALS and ALS-U}

\author{S.D. Murthy\thanks{sdmurthy@lbl.gov}, A. Jurado, M. Betz, Q. Du, B. Flugstad, LBNL, Berkeley, CA 94720, USA}

\maketitle

\begin{abstract}
The ongoing work to upgrade ALS to ALS-U demands strict RF requirements such as low jitter and low spurs frequency reference to meet its accelerator and science goals. A low phase noise dual frequency Master Oscillator (MO), where the two frequencies are related by a fractional ratio of 608/609 and flexible divide by four frequency outputs has been consolidated into a single chassis. Optical fiber clock distribution system has been selected over the old coax system used in ALS to distribute these signals to various clients across the facility, providing high electrical isolation between outputs and therefore lower phase errors. A Xilinx FPGA ties the MO chassis together by providing a RS-485 interface to monitor and control the system. The new system aims to deliver phase continuous frequencies with a phase noise (integrated RMS jitter) from 1\thinspace Hz to 1\thinspace MHz of less than 200\thinspace femtosecond per output. This paper will discuss the design, implementation, performance and installation of the new MO generation and distribution system.
\end{abstract}

\section{INTRODUCTION}
Advanced Light Source Upgrade (ALS-U) is an ongoing project that will upgrade the current ALS system and promises to deliver orders of magnitude increase in brightness and flux of 1\thinspace keV soft X-rays at the diffraction limit. The ALS-U project involves the addition of new 2.0\thinspace GeV Storage Ring (SR) in existing tunnel optimized for low emittance, and a new 2.0\thinspace GeV Accumulator Ring (AR) for full energy swap-out injection and bunch train recovery. 

The legacy MO system at the ALS dating back to 1989 is an encapsulation of crystal oscillators with distribution amplifiers to provide a reference at $f_1$ $\approx$ 499.64\thinspace MHz to the entire machine through phase-stable coaxial cables \cite{als}. An AD9513 evaluation board was employed to generate an additional divide by four clocks required for the timing, gun, and buncher LLRF systems, and a local multiplier was used to provide the $6\times$$f_1$ frequency needed by the S-band linacs. Most of these original parts have become obsolete and in recent years the the beamline experimental setups now require lower phase noise and are more sensitive to spurs. In addition to obsolete parts, the existing distribution chassis was found to add a significant amount of phase noise to the signal.

\begin{figure}[!htb]
   \centering
   \includegraphics*[width=0.85\columnwidth]{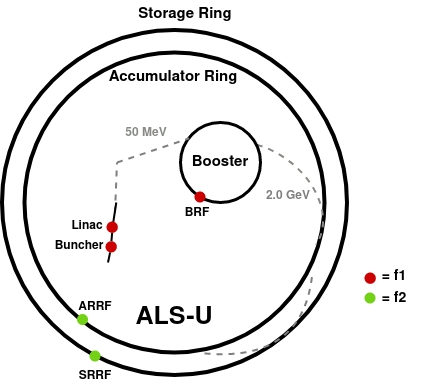}
   \caption{RF systems at the ALS-U, with the reference frequencies.}
   \label{fig:setup}
\end{figure}

The installation of the new SR and AR will increase the MO RF frequency to $f_2$ $\approx$ 500.39\thinspace MHz, and a choice was made to operate them along with the beamline at this frequency while keeping the linac and booster at the new lower frequency $f_1$ = $\frac{608}{609}$$\cdot$$f_2$ $\approx$ 499.56\thinspace MHz \cite{internal1}. The ALS-U layout with frequency references, as currently planned, is shown in  Fig.~\ref{fig:setup}. 

The current coax based MO distribution system is not flexible to support this new dual frequency configuration and will be replaced by an optical fiber based system. The advantages to the optical distribution system include excellent EMI and crosstalk rejection, inherent galvanic isolation, significantly low signal loss, higher fan-out capability, and the absence of powerful microwave distribution amplifiers with their potential power dissipation and signal quality issues \cite{wiki1}.

\begin{figure}[!htb]
   \centering
   \includegraphics*[width=1.0\columnwidth]{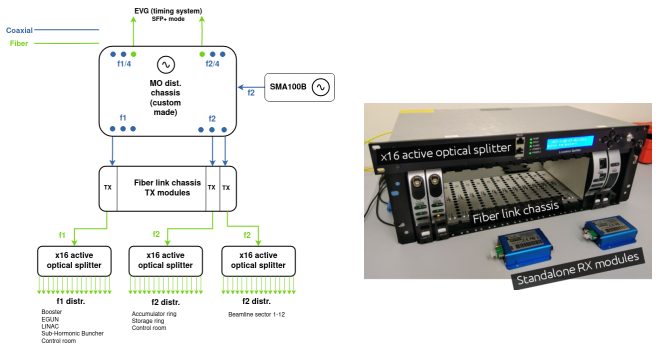}
   \caption{ Left: Layout of the new fiber based MO distribution system for the ALS-U. Right: Vialite components which will be used.}
   \label{fig:fiber}
\end{figure}

The commercial RF over Fiber (RFoF) system from Vialite has been chosen due to its very good phase noise performance. This includes fiber link chassis to convert RF into optical signal and then active optical splitters to distribute them as shown in Fig.~\ref{fig:fiber}.

The requirements for ALS-U MO include the lowest possible phase noise and low spurious coherent signals (spurs) in the 1\thinspace Hz to 100\thinspace kHz offset range. The spurs in this range will transfer through the LLRF system onto the beam, modulating the bunch arrival time. For a slow orbit feedback system, frequency setpoint needs to be adjustable in 1\thinspace Hz increments around $f_2$ \textpm 10\thinspace kHz with $\sim$1\thinspace Hz update rate. Output phase and amplitude need to be continuous during these adjustments, which in turn applies to $f_1$ outputs as well.

\section{System Design}
The proposed final ALS-U MO system consists of commercial signal generator (Rohde and Schwarz SMA100B) to generate the new $f_2$ frequency, followed by a custom-built distribution chassis to generate $f_1$, $\frac{f_1}{4}$ and, $\frac{f_2}{4}$ along with the optical fiber system \cite{llrf2019}.

\subsection{Hardware}
\begin{figure}[!htb]
   \centering
   \includegraphics*[width=0.9\columnwidth]{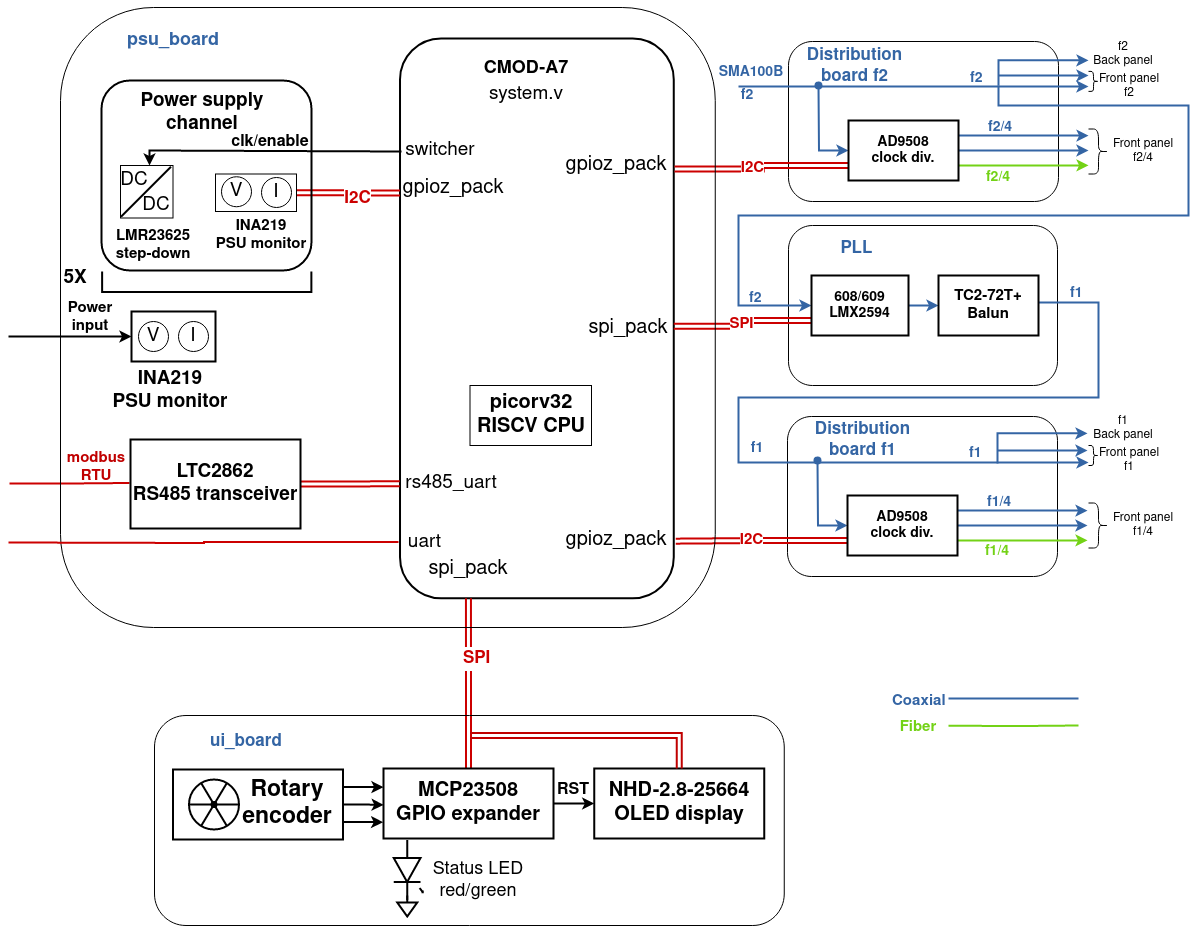}
   \caption{Logical diagram of MO distribution chassis.}
   \label{fig:block_diagram}
\end{figure}

\begin{figure}[!htb]
   \centering
   \includegraphics*[width=0.85\columnwidth]{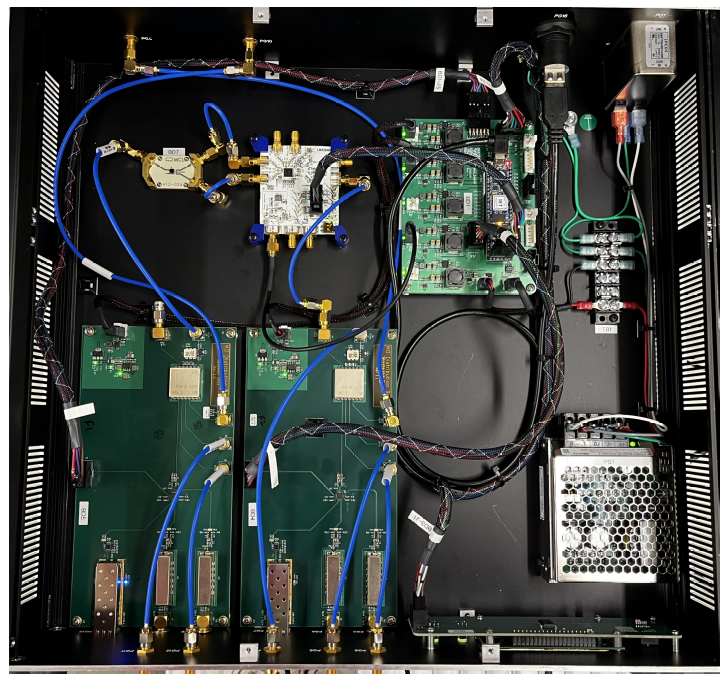}
   \caption{Production MO distribution chassis.}
   \label{fig:pict}
\end{figure}

The new distribution chassis consists of two distribution boards - one for each $f_1$ and $f_2$ frequencies, an LMX2594 PLL evaluation board - to generate  $f_1$ = $\frac{608}{609}$$\cdot$$f_2$, a Power Supply Unit (PSU) - for power management, control UART and Modbus communication, and frequency counters, a CMOD-A7 FPGA module, and User Interface (UI) board - user interface and display as shown in Fig.~\ref{fig:block_diagram}. The production MO distribution chassis is shown in Fig.~\ref{fig:pict}.

\begin{figure}[!htb]
   \centering
   \includegraphics*[width=0.85\columnwidth]{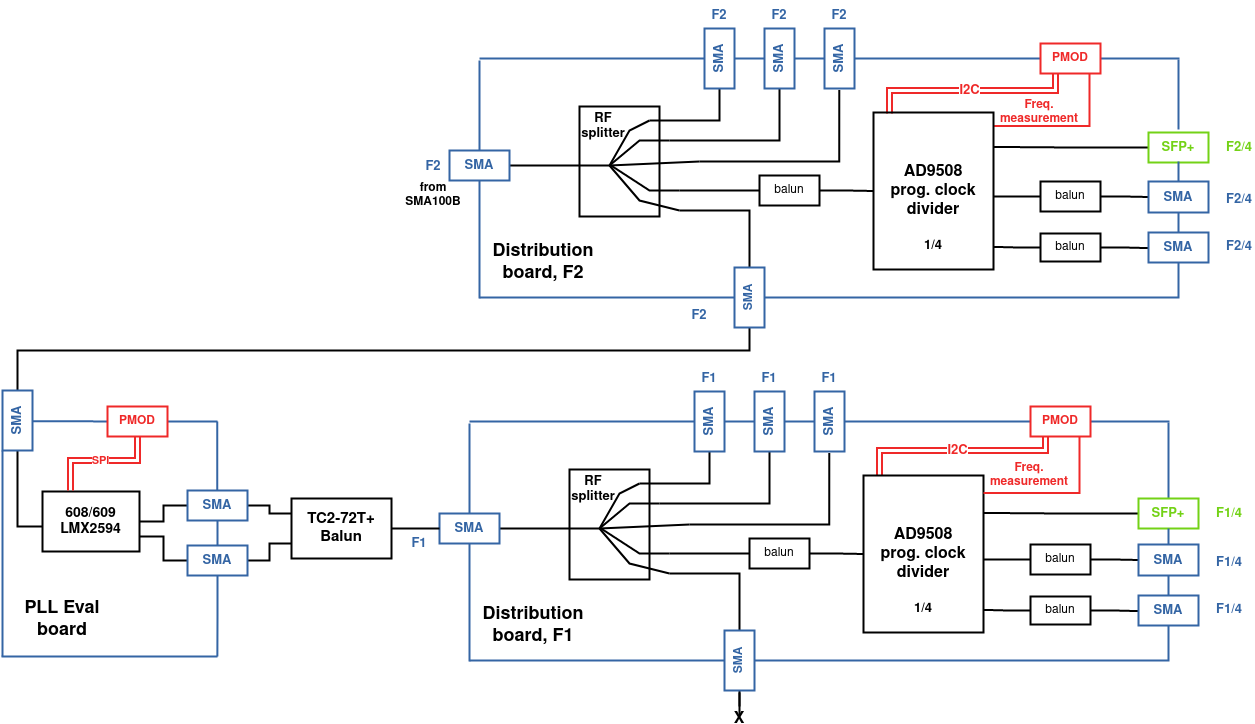}
   \caption{Hardware architecture with signal flow.}
   \label{fig:analog}
\end{figure}

The two distribution boards are identical and use AD9508, a four channel programmable clock divider to divide $f_1$ and $f_2$ by four. One of the output channels is used for frequency measurement in the FPGA and the rest of them are distributed to various clients through Vialite rack/splitter. One of the three outputs generated is in the optical domain and is directly connected to the timing system, while the rest of them are on coax. The chip also provides automatic synchronization of all outputs therefore they all start with the same phase as shown in Fig.~\ref{fig:analog}.
The second frequency $f_1$ is generated using a fractional PLL LMX2954 capable of achieving very low in-band noise, integrated jitter, and reduced spurs. It can turn off output ($f_1$) when not locked and also provides dynamic amplitude level control.

\subsection{Firmware and software}

A small form factor CMOD-A7 board built on Xilinx Artix-7 FPGA with a RISC-V capability ties the distribution chassis together. An open-source, size-optimized RISC-V CPU PicoRV32 \cite{wiki2} is used to handle system configuration, boot-time self-checking, continuous status monitoring, and remote interfacing with EPICS via the RS485/Modbus RTU port. The footprint of the firmware is only 60 kB of RAM and 22\% LUTs are used.
The firmware provides serial communication protocols through the PMOD connectors such as SPI to communicate to the PLL and the UI boards and I2C to communicate to the AD9508 clock divider chip. It is also equiped to measure frequencies $f_1$, $f_2$ , $\frac{f_1}{4}$ and, $\frac{f_2}{4}$ through counters with an accuracy of about 500\thinspace Hz.

\begin{figure}[!htb]
   \centering
   \includegraphics*[width=0.9\columnwidth]{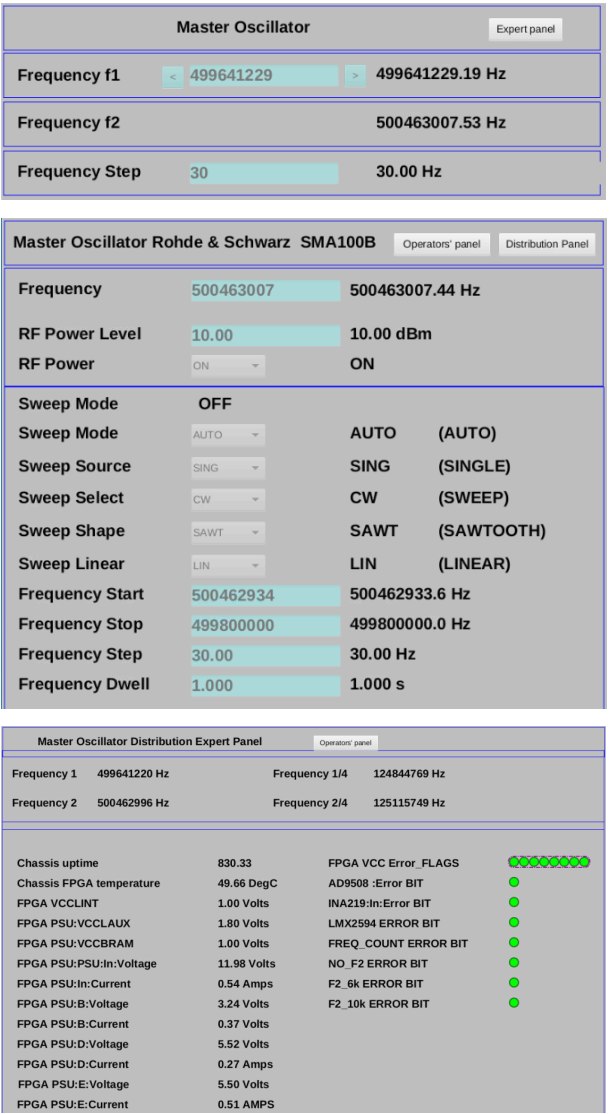}
   \caption{ Top: SMA100B operator panel. Center: SMA100B expert panel. Bottom: Distribution chassis panel.}
   \label{fig:epics}
\end{figure}

The host server is continuously polling the registers through the interrupt service function of the PicoRV32 CPU, which handles Modbus RTU protocol over Ethernet and retrieves real-time register information for monitoring. Some of the available registers include voltage and current monitoring, continuous error detection (communication error, PLL not locked, and $f_2$ frequency drifts), and frequency counters. The UI board also displays these status registers through the OLED screen. An EPICS IOC is built based on this register mapping and a Phoebus engineering screen was developed to display both the MO and the distribution chassis status at both operator and expert level panels as shown in Fig.~\ref{fig:epics}. The software also promises to archive the frequency data available for long-term study.

\section{Installation and Commissioning}
The new optical fiber based MO system along with the distribution chassis is going to be installed in three stages over the course of the next few years: the Pre-Dark Time - divided into two stages, before AR and during the AR commissioning periods and Post-Dark Time - final ALS-U configuration in 2025. 

\begin{figure}[!htb]
   \centering
   \includegraphics*[width=1.0\columnwidth]{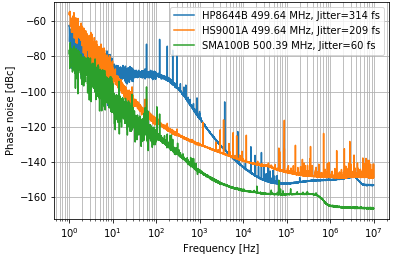}
   \caption{Phase noise comparison plots of different commercial MO.}
   \label{fig:phase_noise_mo}
\end{figure}

In preparation for this, the first step was taken to change the hierarchical crystal oscillator source to a commercial HP8644B signal generator which reduced the overall phase noise to 314\thinspace fs within 1\thinspace Hz to 10\thinspace MHz frequency range. In January 2019, the HP8644b source was replaced by Holzworth HS9001A to reduce the integrated phase noise to 0
209\thinspace fs and avoiding the usage of external DC control voltage for frequency tuning. Finally, in December 2022 the HS9001A was switched to an ultra low-noise Rohde and Schwarz SMA100B, thereby further reducing phase noise to 60\thinspace fs (5X better compared to HP8644B) as shown in Fig.~\ref{fig:phase_noise_mo}.

The distribution chassis, Vialite fiber link chassis, and the three Vialite splitters were installed during the September 2023 summer shutdown. The distribution chassis currently has two clients: $f_1$ coax output - to the entire system through the old MO distribution system and the $\frac{1}{4}$ divider, $f_1$ fiber output - to the Sub-Harmonic Buncher (SHB) through the vialite system. Further down the commissioning phase the old distribution system with the $\frac{1}{4}$ divider will be retired and the new system will drive the entire accelerator. 

\section{Performance}
\begin{figure}[!htb]
   \centering
   \includegraphics*[width=1.0\columnwidth]{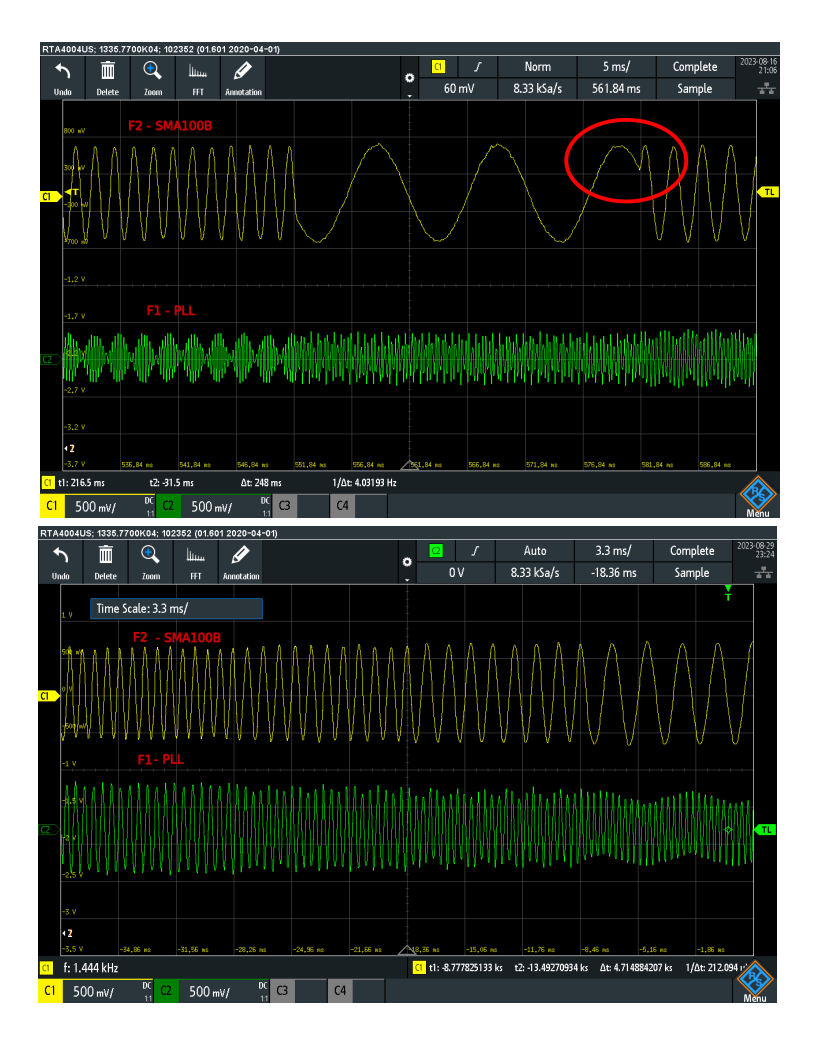}
   \caption{ Top: SMA100B $f_2$ and LMX2594 PLL $f_1$ outputs are discontinuous for step size 500\thinspace Hz. Bottom: Continuous outputs from both SMA100B and PLL with step size of 300\thinspace Hz.}
   \label{fig:cont}
\end{figure}

The phase and amplitude continuity tests were first performed with the new production distribution chassis and the SMA100B. There was no glitch in LMX2594 PLL outputs as long as the SMA100B did not have phase jumps or signal dropout while changing/sweeping frequencies. This can be achieved by limiting the SMA100B's step size to < 300\thinspace Hz during the sweep process. Fig.~\ref{fig:cont} demonstrates this feature, with the step size of 500 Hz (top waveform) the phase of SMA100B output ($f_2$) is discontinuous but with a step size of 300 Hz (bottom waveform), there is no glitch in the outputs of both SMA100B and LMX2594 PLL.

\begin{figure}[!htb]
   \centering
   \includegraphics*[width=1.0\columnwidth]{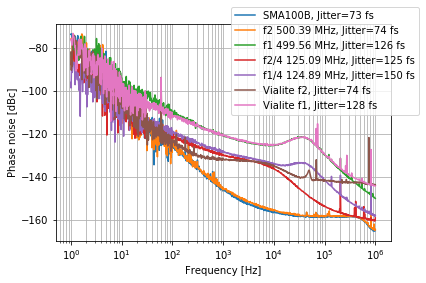}
   \caption{Measured phase noise and spurs of the new MO system at every output stage.}
   \label{fig:phs_noise}
\end{figure}

Fig.~\ref{fig:phs_noise} shows the integrated phase noise of the new system in 1\thinspace Hz to 1\thinspace MHz offset range measured using Rohde and Schwarz FSWP phase noise analyzer with RBW of 2\%. The blue trace is the phase noise of the SMA100B output with the phase noise of 73\thinspace fs and the following four traces are phase noise plots at different outputs of distribution chassis, with the measured jitter of $f_2$ = 74\thinspace fs, $f_1$ = 126\thinspace fs, $\frac{f_2}{4}$ = 125\thinspace fs, and $\frac{f_1}{4}$ = 150\thinspace fs. The next two traces are the output of the Vialite splitters at $f_2$ and $f_1$ and show they add no significant noise to the system. As the requirements indicate, the outputs exhibit no significant spurs in the 1\thinspace Hz to 100\thinspace kHz offset range.

\section{CONCLUSION}
The new dual frequency MO system based on the single chassis design and fiber technology was successfully installed and is currently driving the entire system at ALS. More clients will use this new configuration further down the AR commissioning schedule. The system successfully meets the performance requirements of phase and amplitude continuous frequency outputs with the measured phase noise from 1\thinspace Hz to 1\thinspace MHz of less than 200\thinspace fs per channel. Further optimization of the PLL loop filter can be easily achieved if requirements change during the final ALS-U commissioning.

\section{ACKNOWLEDGMENTS}
This work was supported by the ALS and ALS-U Projects and the Office of Science, Office
of Basic Energy Sciences, of the U.S. Department of Energy under Contract No. DE-AC02-05CH11231.
%
%
\ifboolexpr{bool{jacowbiblatex}}%
	{\printbibliography}%
	{%

	
} 
%
%


\end{document}